# What Does Information Science Offer for Data Science Research?: A Review of Data and Information Ethics Literature

Brady D. Lund and Ting Wang

**Abstract**

This paper reviews literature pertaining to the development of data science as a discipline, current issues with data bias and ethics, and the role that the discipline of information science may play in addressing these concerns in data science. Information science research and researchers have much to offer for data science, owing to their background as transdisciplinary scholars who apply human-centered and social-behavioral perspectives to issues within natural science disciplines. Information science researchers have already contributed to a humanistic approach to data science ethics within the literature and an emphasis on data science within information schools all but ensures that this literature will continue to grow in coming decades. This review article serves as a reference for the history, current progress, and potential future directions of data science and ethics research within the corpus of information science literature.

# What Does Information Science Offer for Data Science Research?: A Review of Data and Information Ethics Literature

Data as a subject and data science as a field of inquiry have received immense attention in the past handful of years, as big data becomes more ubiquitous and the applications of machine learning or artificial intelligence (AI) algorithms to process this data become more refined. Information science researchers working in Information Schools (iSchools) around the world have welcomed this development, integrating data-related curricula in their programs and encouraging data science research among faculty (Ortiz-Repiso et al., 2018; Song & Zhu, 2017; Wang & Lin, 2019). Meanwhile, the tremendous development of data storage capabilities due to the automation of measurement and data collection procedures, and the creation of highly sophisticated tools for analyzing and processing data, have given rise to considerable philosophical and legal concerns about the correct, legal, and appropriate approach to use data (Hand, 2018).

On the one hand, information science is directly concerned with the storage and management phases of the data life cycle (Marchionini, 2016). On the other hand, one area of particular importance within the library and information science (LIS) is ethical practices: professional codes of ethics, anti-discrimination, and freedom from bias (Rubin, 2017). The human side of data science, the impact of how data is impacted by and can impact people, may then be seen as one of the areas where information science researchers and professionals can contribute expertise to the key discussions within the field of data science. Humanistic data science has already been seen in a few of the highly-notable works published by LIS researchers in recent years (Floridi, 2021; Hoffman, 2019; Noble, 2018), which may contribute to the future development of data science. However, there is a lack of understanding, at present, of how information science researchers have uniquely contributed to a discussion of data ethics. This paper will work to address these gaps by reviewing the development of the literature pertaining to the conception of data ethics and data bias and the movement towards a more humanistic data science within the information science literature.

The relevant literature for this review was retrieved within the *Library and Information Science Source* (Ebsco), *Computer Source* (Ebsco), and Google Scholar databases from July 2021 to April 2022. Each of the retrieved papers was reviewed by the authors to determine its relevance to the review and importance for supporting the complete narrative of the development of data ethics within the information science literature. This paper is designed as a narrative review (Paré et al., 2015), meaning that the articles selection aims to illustrate the key findings/concepts in relation to the central questions rather than necessarily provide a scoping review of *all* literature available.

Selected articles were chosen with the aim of addressing the following questions:

- *How do data bias and the evolution of data ethics appear in information science literature?*

- *What perspectives on data science bias and ethics within information science appear in the literature?*

This study covered the overlap of data science and information science, data bias and data ethics, and information bias and information ethics according to the findings of the collected literature to explore the possibility of the relatively young data science promoting its future development from the perspective of humanity science by learning from information science. For the purpose of this study, "information science" is used to mean the academic transdiscipline which studies all aspects of information (Taylor, 1963; Borko, 1968), whereas "library and information science" is used to refer both to the discipline of information science and professional training and activities of librarianship (library science) that are informed by this discipline (Bates, 1999; Butler, 1951).

**Overlap of Data Science with Information Science**

The term "data" is used within the field of data science usually with the concept of electronic data or information, which has evolved over several centuries through the work of individuals like George Boole (1847), Alexander Graham Bell (1881), Ralph Hartley (1928), Alan Turing (1936), and Claude Shannon (1948), in the fields of mathematics, physics, electronics, communications, and computer science. In their work, data is the basic unit that makes electronic communications and devices like the computer possible, but there is also a basis for viewing data as a commodity that can be owned, traded, and used for prediction, and modeling, which is the basis of the modern boom in data science and engineering (Blum et al., 2020; Dhar, 2013; Liang et al., 2018).

Data science emerged from a variety of disciplines over the course of half a century. Perhaps the most direct and prominent influence was from statistics, where papers dating well back into the 1960s discussed the process of collecting, storing, cleaning, and analyzing data (Ball & Hall, 1967; Horst, 1965; Tukey, 1962). Elements of what would become the field of data science are also evident within the history of bibliometrics. Eugene Garfield discusses, in his writings, bibliometrics as the use of citation *data* to "write the history of science" (Garfield et al., 1964). The natural sciences also played a significant role in the development of data science, by helping researchers understand the importance of data contexts throughout the lifecycle of data (Spector, 1956; Coombs, 1964). With the expansion of data collection and availability of data sets in the Internet age, data science has emerged as a distinct field of study (Cleveland, 2007).

Provost and Fawcett (2013) define data science as "a set of fundamental principles that support and guide the principled extraction of information and knowledge from data" (p. 52) with noting that, "at the moment, trying to define the boundaries of data science precisely is not of the foremost importance," as the definitions are constantly evolving as the field grows and refines itself (Provost & Fawcett, 2013, p. 52). One element of the definition of data science that is clear is that it focuses broadly on changes in states of matter within the DIKW (data, information, knowledge, and wisdom) hierarchy (shown in Figure 1 below): transforming data to create information and disseminating findings to create new knowledge (Rowley, 2007). Authors have noted the considerable overlap between the modern discipline of data science and other disciplines, including informatics, statistics, computer science, and information science (Brennan et al., 2018; Furner, 2015; Marchionini, 2016; Zhang & Benjamin, 2007).

**Figure 1. Depiction of DIKW Pyramid**

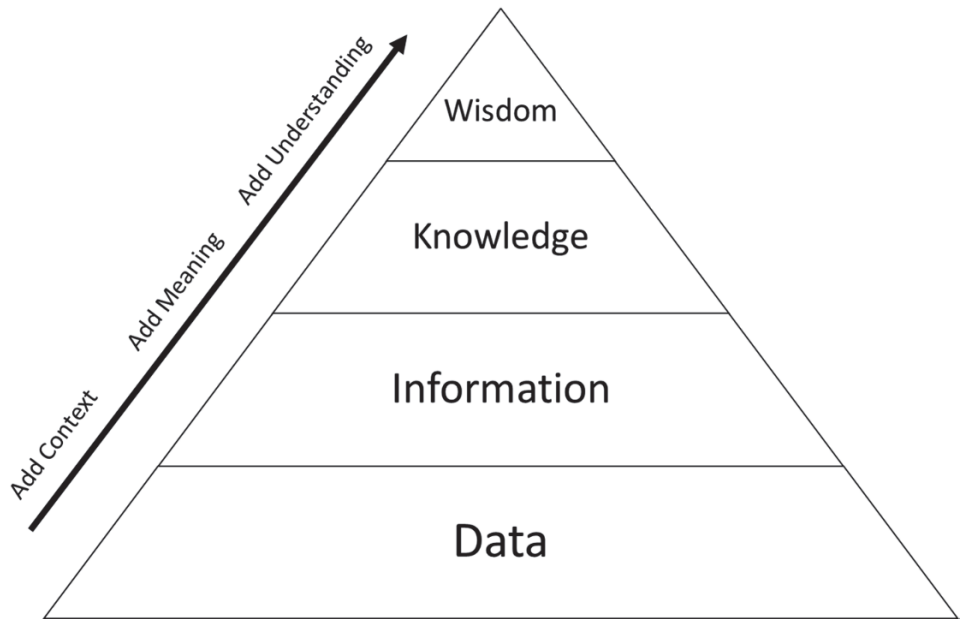

Among library and information science researchers, there was some early interest in data analytics in the period stretching from the 1940s to the 1960s (Batts, 1966; Janeway, 1944; Nicholson & Bartlett, 1962). However, more substantial interest appears to have emerged in the 1970s, with libraries at research universities serving as early repositories for research data, along with a growing interest within the discipline in bibliometrics (Conger, 1976; Pope, 1975; White, 1977). With the growth of information science elements within the discipline in the 1980s, terms like "data and information," "data processing," and "data analytics" become more ubiquitous in the literature (Cuadra, 1982; Griffiths & King, 1985; Wallace, 1985). In succeeding years, the work of bibliometricians carried the growth of the data science elements within information science up until the current era of enhanced interest in the topic among all factions of the discipline during the middle-to-late 2010s (Bar-Ilan, 2008; Egghe & Rousseau, 1990; Thelwall, 2004).

Data science has grown in interest at a rapid rate in recent years, including within the discipline of information science. Just in the past five years, the number of articles relating to data topics published in information science journals has grown by nearly eight-fold (Virkus & Garoufallou, 2019; Ma and Lund, 2020; 2021). Ongoing efforts are pushing for a great intwining of data science with information science, as evident in the work of authors like Wang (2018), Poole (2021), Washington Durr (2020), and Chohdary et al. (2021). While several studies have documented the increasing interest in this topic, few have attempted to parse out what, if anything, makes different disciplinary approaches towards data science unique.

**Data Bias and the Evolution of Data Ethics**

Bias in the analysis of data has long been a concern of analysts, though, until recent years, the focus was mainly on issues with the collection of data, the bias of the analysts themselves, and the imperfection of statistical models to reveal desired findings (Haussler, 1988; Kiviet, 1995; Mark et al., 1999). For instance, Dolly and Tillman (1974) provide a review of early instances of data bias owing to the improper design, collection, or analysis of data in educational research. Klein (1953), along with other researchers in the area of econometrics, also made early steps toward criticizing the quality of data and the impact of poor data on prediction and modeling. Many of these instances of bias were due to a lack of proper consideration of the attributes of a population, leading to a skewed sample.

Only in more recent years, with the emergence of increasingly large datasets and automated methods of analysis, have a range of pressing new concerns arise. A growing interest in artificial intelligence and machine learning, using data about the public to inform decision-making and model trends, has provoked this heightened interest (Koene, 2017; Ntoutsi et al., 2020). Algorithmic bias, flaws in the design of machine learning algorithms that are used to inform decision-making or the quality of the data that is fed to these algorithms, gained traction as a topic of substantial research interest in the mid-2010s, with the work of researchers like Bozdag (2013), Hajian et al. (2016), and Noble (2018). Similarly, the implementations of large-scale datasets to drive artificial intelligence and smart technologies can be deeply flawed when the person interacting with the AI technology fits outside the "norm" or training set utilized (Whittaker et al., 2019).

There are many types of data bias that have been discussed in the literature in recent years. Here, five of these types are explored: sampling bias, availability bias, methodological bias, contextual bias, and interpretation bias. Each of these types of bias may be known by slightly different names depending on the source (e.g., sampling bias may be called "selection bias" by some authors). In order to understand when and where each type of bias occurs, it is useful to reference a model of the data life cycle. The life cycle presented in Figure 2 is a simple model adapted from Stobierski (2021) at the Harvard Business School.

**Figure 2. Data Life Cycle Model**

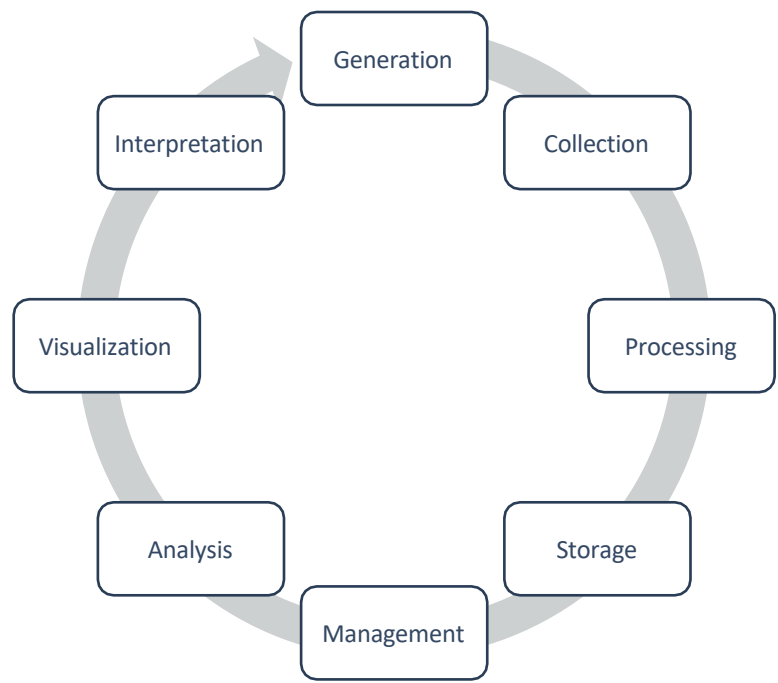

The first phase of the data life cycle is "generation," or the process of either creating or determining the qualities and quantities of data that will be collected (Ray, 2014). Data is then collected and processed (refined and organized for storage), and stored within a database, using some management processes to ensure *contextual integrity* (Nissenbaum, 2004). Next, the data is retrieved for analysis (producing some insight from the data – aka information), visualization (depictions of significant findings), and interpretation (the significance/meaning of the findings). The life cycle of data is continuous, as the data may be reused, archived, or partially or fully deleted, as is depicted as a distinct step in some other models of the data life cycle (Rüegg et al., 2014).

Sampling Bias occurs during the planning and/or collection phase of the data lifecycle (Hahn et al., 2005). Sampling bias occurs when the parameters used to select a sample are improperly designed or executed (Smith & Noble, 2014). This bias can occur due to ethical, logical, and practical faults. Ethical faults occur when the sample collection is skewed due to some limitation owed to ethical barriers, such as a risk to certain key members of a group that prevents participation (e.g., certain victims of domestic violence) (Dattalo, 2010). Logical faults occur on the part of the data collectors when they do not select a truly representative sample (Martinez-Mesa et al., 2016). This has historically been a major, and highly visible, issue with political polling (Lynn & Jowell, 1996). An example of a practical fault is found in the work of Severson and Ary (1983), who reported sampling bias due to consent procedures with minors. Certain parents were simply unwilling to allow their child to participate in a study, and this prevented a representative sample from being collected.

Availability Bias may be classified as a type of sampling bias, which is particularly discriminatory against heterogeneous populations. It occurs when samples are produced and interpretations/decisions are made on the basis of those subjects/data that are most readily available and abundant (Dube-Rioux & Russo, 1988). The impact of availability bias is variable upon the purpose for which the data was collected. In medical facilities, availability may lead to the misdiagnosis of a rare medical condition as one that is more common (Mamede et al., 2010). In economics, it may lead to an investor selecting a stock that is widely known or popular over one that is likely to produce the best return (Javed et al., 2017). In library and information science, it may result in the accumulation of library materials from a few popular, local, or cheap sources over those that offer the most diverse and intellectually challenging materials (Harmeyer, 1995).

Statistical Bias is the product of the statistical analyses selected and their execution. There is, of course, a bias that is inherent to statistical tests (Cole, 1981; Millsap & Everson, 1993). There is also bias owing to the sample characteristics of the data relative to tests selected – skewed data, outliers – that can lead to incomplete or improper analyses and interpretation (Chen et al., 2001; Kinksman, 1957; Westgard & Hunt, 1973).

Context Bias refers to bias that occurs when the context from which data was collected or analyzed is lost during the storage, analysis, or interpretation stages of the life cycle. In some cases, contextual bias can actually be an un-bias of sorts, such as in the forensic sciences where knowing information about how something originated could negatively influence interpretations of events (Osborne et al., 2014). However, in most cases, removing context from a situation is very problematic. It leads to interpretations of a situation – such as in the case of medical diagnosis or investment strategy – that may be accurate in some cases but vastly inaccurate it others (Price & Nicholson, 2019).

Interpretation Bias occurs when the interpretation or ways in which a story is told about data findings is itself biased (Wilholt, 2009). Interpretation of data can be influenced by one's past experiences and cultural biases, such that a person from one background may interpret the findings of analysis very differently from those from another background (Lund, 2022). All research publications rely on interpretations of some kind of data (Gummesson, 2003; Šimundić, 2013). Though some have noted strategies for mitigating this type of bias, such as using a blinded interpretation of findings by individuals not directly involved with the data collection and management (Järvinen et al., 2014), these are still not perfect defenses against the effect of bias.

As recognition of data bias, algorithmic bias, and AI malpractice grew over the past several years, so too did the promotion of new ethical standards and education for data scientists. Mittelstadt and Floridi (2015) and Barocas and Boyd (2017) survey current ethical beliefs and practices of data scientists and identify gaps that may exacerbate concerns with the next generation of machine learning technologies. Several authors have proposed some model or overarching philosophy for the responsible use of big data analytics (Fairfield & Shtein, 2014; Richards & King, 2014; Zwitter, 2014).

One of the distinguishing elements of information science research from systems or computer science research is its emphasis on the *ethical* use of information/data, rather than primarily emphasizing only how best to streamline its use. Information science has a long history of debating the ethical underpinnings of what has evolved into the fields of data science/mining and artificial intelligence (Adam, 1991; Carlin, 2003; Kostrewski & Oppenheim, 1979; Shiri, 2016). This approach is one major benefit of information science's interdisciplinary nature and social science bent. The discipline's foundations within the profession of librarianship, and its codes of ethics, strongly influence the current philosophy toward data topics (Roeschley & Khader, 2020; Shachaf, 2005). **Perspectives on Data Science Bias and Ethics Within Information Science**

This section provides an overview of recent work within the literature of information science that suggests an interest in developing ethics for data science. Interest in this area of study is still growing (Ajibade & Mutula, 2020; Ma & Lund, 2021), but already a sizeable collection of research and essays exists on this topic, dating back over the past decade.

Shilton (2012) discusses the ethics of participatory personal data (PPD) from an information science perspective. PPD is "a new class of data about people and their environments, generated by an emerging network of embedded devices and specialized software" (Shilton, 2012, p. 1906). It includes data captured by users on personal devices, like geotagging on photos, GPS tracking on a smartwatch or other exercise device, and "smart" sensors and devices. This data clearly presents many significant ethical challenges within the purview of information science researchers and professionals:

- Surveillance concerns (who has access to this data, which could be used to reveal confidential insights about the individual),
- Storage of data (ensuring that it is not compromised), the use of this data for research purposes,
- Access to these devices and usage of personal data by the users themselves.

Andres (2016) analyzes the ethics of competitive intelligence (CI) through the lens of information science. These fields, traditionally, take starkly different tactics in approaching how data/information should be used. While CI is perhaps not as ruthless as "competitive," the name seems to imply the beliefs about how and why data may be used differ from the standards that library and information professionals apply in utilizing and sharing information in their work. One major distinction stem from the data collection methods, for instance. Library and information professionals are taught to consider awareness and consent vital in data collection (Cooke, 2018; Rubel, 2014). These elements are important in CI as well, but definitions of "awareness" and "consent" may differ. While many entities that collect and utilize users' data today employ those lengthy consent forms that research shows that most users will never fully read or understand (Cassileth et al., 2018; Pederson et al., 2011), library and information professionals might opt for a more succinct and clear description of how the data will be collected/used (in addition to the lengthy legal form). People can often be a burden for the data analyst – but not for information professionals, who place the human being at the center of their work.

Marchionini (2016) argues that information science is one of four foundational pillars of data science (along with statistics, computer science, and knowledge domains) and can be seen as lending theory (which extends to ethical practices) to the emerging field. Information science is particularly important in terms of data curation, which is influenced by principles of knowledge organization, modeling, and storytelling that have been the expertise of library and information professionals for over a century. Information scientists, according to Marchioni, are unique in their focus on information throughout its lifecycle, including "ethical and legal conditions associated with data collection," "appraisal, data quality, and cleaning," "metadata assignment," "storage and preservation of data," and "evaluation of conclusions based on data exploration and analysis and making the data and workflows findable and reusable" (Marchionini, 2016, p. 5).

Several articles published in recent years have begun to push information science towards a more defined role within data science. Hoffmann et al.'s (2018) panel discussion at the ASIST annual meeting ignited elevated exchange on the topic, which had already been brewing somewhat within the discourse of the field of IS. Each of the presenters of this panel hasengaged in scholarship that accelerates debate on topics like fairness and discrimination with AI (Hoffmann, 2019; Roberts, 2016; Wood, 2017).

One significant contributor to these discussions about data ethics from an information science perspective is Luciano Floridi. Floridi's work (e.g., Floridi, 2014; Mittelstadt et al., 2016) has received great recognition and acclaim among information science researchers, including Fyffe (2015), Wang (2018), and Bawden and Robinson (2020). Among the many important ideas lent by Floridi to information science is an ethical philosophy of data and artificial intelligence. One concern with AI is not that it poses a legitimate threat to human livelihood but that its benefits for human quality of life will be disproportionately felt by certain populations and exacerbate disparities. These disparities can already be seen in the national investment in AI if one looks at the investment of the United States and China relative to most nations in the Global South, which have limited funds to invest in these endeavors.

Within library and information science, it is understood that information should be available to all. Ranganthan's (1931) second law of library science states, "Every person his or her book" and "BOOKS ARE FOR ALL" (p. 74) – access is built into the foundation of the discipline of information science. However, access is traditional and not a guiding principle in data analytics. Data is highly valuable and is generally held in such a way, along with the technology and intellectual capital necessary to store and analyze the data (Banterle, 2019; Ballantyne, 2020). This exacerbates the class divide and further eliminates the potential for economic mobility (Rolfe, 2017; Schradie, 2020). An information science-inspired effort to expand access to the tools and technologies necessary to compete in a data-driven economy is intriguing, to say the least. But how could such an endeavor be accomplished? Already the information access mission of libraries largely fails when the picture is expanded to include countries in the Global South, where library resources and technology are far lesser than in the Global North (Lund et al., 2021; Pather & Gomez, 2010).

Roeschley and Khader (2020) provide the most extensive review of data ethics to date in their iConference paper. In their study, the researchers reviewed data science articles published in library and information science journals, identifying four themes relating to data ethics issues addressed in the literature: privacy, research ethics, ethical ecosystems, and control. The researchers note that the first theme, privacy, dates back to the earliest values of the field, set forth in the founding documents of organizations like the American Library Association and the Association for Information Science and Technology (as noted previously in this review). Research ethics, similarly, has a long history within information science, with academic librarians' long-held role in the lifecycle of scholarly research. The ethical handling and role in the lifecycle of scholarly research. The ethical handling and reporting of research data is anticipated or presently considered by many information science researchers to be a central role of the modern academic library (Cox & Pinfield, 2014; Laskowski, 2021; Pinfield et al., 2021).According to Roeschley and Khader (2020, p. 3), ethical ecosystems are "environments in which one would expect to see a common standard for ethical behavior." The researchers note that these environments may include business, academic, and government environments. Certainly, this role aligns with the traditional roles of libraries as complex information ecosystems, guided by a shared set of ethical standards (Walter, 2008). Lastly, control relates to maintaining the data security and understanding the legal limitations pertaining to the data (Roeschley & Khader, 2020). The term "control" conjures to mind the subject of knowledge organization, bibliographic control, and the management of metadata. Indeed, the two roles (cataloger and data manager) can be seen as quite similar. The work of Roeschley and Khader (2020) provides a foundation for future study, which may explore more deeply the unique contributions of information science research and researchers to anti-bias and ethics literature in data science.

## Discussion

Growth in data science research and data professionals over the coming decades is a near-certainty. The value of data to both the profit-generating and non-profit sectors of the economy are undeniable and will assuredly drive investments within higher education and among grant-funding organizations (Berman et al., 2018; Cao, 2017; Waller & Fawcett, 2013). It is not surprising that information schools would be eager to grab a piece of this field, as data science offers tremendous potential for the kind of attention and growth for which these schools are often overlooked. As this paper details, information science has much to offer to data science and, indeed, has already had an impact on the development of this growing field.

**How Has Information Science Contributed to the Development of Data Science?**

Historically, information science has been directly involved in the data life cycle's storage and management phases. Management of data repositories has been a growing role within academic library services over the past several decades (Antell et al., 2014; Brase & Farquhar, 2011; Read, 2008; Shen, 2015; Si et al., 2013). These roles remain important in the profession of librarianship as well as the information science research as particularly evident in the work of Carol Tenopir, among others (Tenopir et al., 2014; Tenopir et al., 2015; Tenopir et al., 2017; Tenopir et al., 2020). In more recent years, with a greater emphasis on the topics of health informatics and data-driven decision-making for health outcomes, data has taken on a greater role within the discipline (Bath et al., 2005; Chen et al., 2012; Corbett et al., 2014; Kerr et al., 2008; Tremblay, 2016). Growth in the potential role of data, new avenues for AI technology, increased research funding, and the impacts of major crises like the COVID-19 pandemic led to a broader role of this topic area within the discipline (Semeler et al., 2017; Shankar et al., 2020; Siguenza-Guzman et al., 2015; Song & Zhu, 2017; van der Aalst et al., 2017). Research data management, health informatics, and data science implications for information organizations are thus three major areas where information science researchers have contributed to the literature on data science.

Certainly, information science also maintains an intimate connection with data science within their adjacent places in the DIKW pyramid. The transdisciplinary work of Marcia Bates illustrates that the concepts of "data" and "information" cannot be truly divorced from one another. As Bates (2005) notes, data is information that can be sensed but not yet consciously understood or has not yet been rationally organized (one might react to a pinprick, but it is entirely different to understand the significance or reason for that pain). Data may be handled in much the same way as information. We see that the contemporary models of the information life cycle and information management align well with those of data life cycle and data management; indeed, researchers in these areas frequently cite one another's work (Christopherson et al., 2020; Cox & Tan, 2018). Additionally, there is an emerging interest among information literacy experts (e.g., academic librarians) to provide instruction in the related area of data literacy (Koltay, 2017; Prado, 2013; Špiranec et al., 2019).

In Bates's (2015) article "The Information Professions: Knowledge, Memory, Heritage," the author conceives a spectrum of the information disciplines, as they span the traditional domains of arts and humanities, social and behavioral sciences, and natural science and mathematics.

Bates situates data science-related sub-disciplines – data mining, scientometrics, data storage, informatics – firmly within the natural sciences and mathematics end of the spectrum, on the opposite end from library science and museum studies, which lie more firmly within the arts and humanities. However, the concept of the information disciplines as a spectrum illustrates how data science may be viewed among information science researchers. There are no defined borders within a spectrum; there can be bleed-over. This differentiates the education and perspectives of the information scientist from the pure natural scientists or mathematicians. Aspects of information science like information policy, semiotics, and information behavior can inform our perspectives on data science, leading to a more humanistic perspective on the field (Aragon et al., 2016; Kuiler & McNeely, 2020).

**What Contribution Has Information Science Provided to Our Understanding of Data Ethics?**

One place to start when answering this question is with Safiya Noble's (2018) *Algorithms of Oppression*, a book that has been cited over 3000 times by authors from across virtually all disciplines, thus making the scope of its impact very clear. The book highlights how the design of algorithms – all phases of the data and information life cycles – is impacted by the implicit (and explicit) biases held by analysts and developers. Of course, Noble's work did not emerge independent of any prior work within academia. It was informed by "critical information studies and critical race and gender studies" (Noble, 2018, p. 6). This includes work by information science faculty like Bar-Ilan (2007a; 2007b), Furner (2007), Saracevic (1999), Smith (1981), Spink et al. (2001), and Zimmer (2008). Fairness in information retrieval is the central focus of Noble's book, a topic discussed by many other information science researchers beyond those previously cited, including Shah (Gao & Shah, 2020a; Gao & Shah, 2020b) and Jansen (Eastman & Jansen, 2003; Jansen & Schuster, 2011).

Contextualization of data is a key theme in the most impactful data science research published by information science journals. Context is an important component of how information science researchers understand the world (Madden, 2000), whether it is the life cycle of information/data or understanding the threat posed by global inequities. Wang (2018) appears to come to a similar conclusion – albeit not in the same words – when they suggest that the greatest opportunities for information science to contribute to data science are through a conception of data science research that is not purely positivistic and by which the quality of data, not merely the quantity, are questioned. Many of the human-centric theories that form the basis of much information science research today – such as Belkin's (1980) "Anomolous States of Knowledge," which describes how limitations of the human ability influence interaction – have direct applicability to how we might attain a more humanistic data science.

## Conclusion

Data science is a rapidly growing field that represents an industry that is expected to reach a value of over $100 billion in the United States by the end of the 2020s (Acute Market Reports, 2020). That valuation alone would be enough to encourage researchers and educators in any discipline to expand their reach into data science. However, information science researchers and

library and information science education programs are particularly well-situated to work within this field of data science. With a background in the life cycle of information, management strategies, and provision of information, the educational background of information science researchers translates well to handling data throughout its life cycle (Semeler & Pinto, 2020). Information science's transdisciplinary nature prepares researchers well for applying humanistic perspectives to combat data bias and impose an ethical regimen on how data is collected, stored, and used.

This review illustrates that information science researchers have already invested efforts into staking a claim within data science research, not merely by expanding faculties to hire more data science researchers from the natural science disciplines, but by educating and performing research in their own right that addresses key issues within this expanding field. Through continued investment and support, information schools have a considerable role in how data science evolves and what universities continue to thrive during a period of elevated financial scrutiny of educational institutions.